\documentclass{SCGE}
\usepackage{hyperref}
\begin{document}

\begin{picture}(0,0){\rm
\put(0,-20){\makebox[160truemm][l]{\bf {\sanhao\raisebox{2pt}{.}}
Article  {\sanhao\raisebox{1.5pt}{.}}}}}
\put(0,-34){\jiuwuhao {\textcolor[rgb]{0.5,0.5,0.5}{\sf 
}}}
\end{picture}

\def\bm{\boldsymbol}

\def\dl{\displaystyle}
\def\du{\end{document}}
\def\d{{\rm d}}
\def\e{{\rm e}}
\def\i{{\rm i}}

\Year{} %
\Month{} %
\Vol{} 
\No{} 
\BeginPage{1} 
\EndPage{??} 
\AuthorMark{{\rm Yin Z Q}, et al.}  
\AuthorMarkCite{{\rm Yin Z Q}, et al.} 
\DOI{} 

\title{Hybrid opto-mechanical systems with nitrogen-vacancy centers}

\author[1]{YIN ZhangQi}{Corresponding author (email: yinzhangqi@mail.tsinghua.edu.cn)}
\author[2,3]{ZHAO Nan}{Corresponding author (email: nzhao@csrc.ac.cn)}
\author[4,5]{LI TongCang}{Corresponding author (email: tcli@purdue.edu)}

\address[{\rm1}]{Center for Quantum Information, Institute for
Interdisciplinary Information Sciences, Tsinghua University, Beijing
100084, China;}
\address[{\rm2}]{Beijing Computational Science Research Center, Beijing 100084, China;}
\address[{\rm3}]{Synergetic Innovation Center of Quantum Information and Quantum Physics, University of Science and Technology of China, Hefei 230026, China;}
\address[{\rm4}]{Department of Physics and Astronomy and School of Electrical and Computer Engineering, Purdue University, West Lafayette, IN 47907, USA;}
 \address[{\rm5}]{Birck Nanotechnology Center, Purdue University, West Lafayette, IN 47907, USA}

\maketitle \vspace{-3.5mm}{\footnotesize\begin{center} Received Month date, Year; accepted Month date, Year
\end{center}}\vspace*{-5mm}

\begin{center}
\rule{16.5cm}{0.4pt}
\parbox{16.5cm}
{\begin{abstract} 

In this review, we briefly overview recent works on hybrid
(nano) opto-mechanical systems that contain both mechanical
oscillators and diamond nitrogen-vacancy (NV) centers. We
review two different types of mechanical oscillators. The
first one is a clamped mechanical oscillator, such as a
cantilever, with a fixed frequency. The second one is an
optically trapped nano-diamond with a build-in nitrogen-vacancy
center. By coupling mechanical resonators with electron spins,
we can use the spins to control the motion of mechanical
oscillators. For the first setup, we discuss two different
coupling mechanisms which are magnetic coupling and strain
induced coupling. We summarize their applications such as
cooling the mechanical oscillator, generating entanglements
between NV centers, squeezing spin ensembles and et al. For
the second setup, we discuss how to generate quantum superposition
states with magnetic coupling, and realize matter wave
interferometer. We will also review its applications as ultra-sensitive
mass spectrometer. Finally, we discuss new coupling mechanisms
and applications of the field.
\end{abstract}}
\end{center}\vspace*{-0.6cm}

\begin{center}
\parbox{16.5cm}
{\bf\jiuhao Opto-mechanics, Nitrogen-vacancy center, Hybrid system}
\end{center}

\begin{center}
{\PACS{\rm ??????}}
\Cit{~~~???, et al. ???. Sci China-Phys Mech Astron, 2014, 57: 1--6, doi: }
\end{center}

\textwidth=178truemm \textheight=236truemm

\wuhao\vspace*{1.5mm}

\begin{multicols}{2}

  \renewcommand{\baselinestretch}{1.08} \baselineskip
  12.2pt\parindent=10.8pt

  \renewcommand{\thefootnote}

  \section{Introduction}\label{sec:intro}

  In 2012, David Wineland, together with Serge Haroche, won Nobel
  prize "for ground-breaking experimental methods that enable
  measuring and manipulation of individual quantum systems"
  \cite{Nobel2012}. In his lab, the motion and the electron spin of
  trapped ions couple with each other. Therefore, the motional degree
  of freedom of the trapped ions can be coherently manipulated, such
  as the ground state cooling, Fock states generating and detecting
  with unit fidelity and efficiency \cite{TrappedIon1}. Using these
  technologies, David Wineland generated Shr\"odinger's cat states
  with trapped ions \cite{TrappedIon2,Monroe96}, which for many years
  only existed in through experiment.

  The "true" Shr\"odinger's cat state is defined by a microscopic
  two-level system entangling with a macroscopic system.  New
  technology, such as opto-mechanics, is needed to generate quantum superpositions in
  macroscopic systems. Opto-mechanics studies the radiation pressure on
  the mechanical oscillator, and how to use this pressure to manipulate
  the motion of mechanical systems. In 17th century Kepler found the
  radiation pressure of light when he studied the comet tails.
  1n 1909, Einstein studied the statistics of radiation pressure force
  fluctuations on movable mirror \cite{Einstein09}.
  In 1960s, Braginsky studied the role
  of radiation pressure in the context of interferometers, and
  proposed to cool the motionnal temperature of the mirror by radiation
  pressure \cite{Braginski67}.
  In 1969, A. Ashkin observed optical trapping of micron-sized particles
  in liquid \cite{Ashkin70}. Two years later, optical levitation
  of glass spheres by an upward-propagating laser beam in both air and
  vacuum was demonstrated \cite{Ashkin71}.
  The optical tweezers now is widely used in atomic physics, chemistry
  and biology.

  Recent years, we witness great developments of opto-mechanics \cite{optoRev2012,optoRev2013,Liu13}.
  We have achieved the
  quantum ground state of macroscopic mechanical oscillators by
  traditional cryogenic techniques in 2010 \cite{OptoGround10}, or by
  active cooling in 2011 \cite{OptoGround11a,OptoGround11b}. We have
  developed many technologies based on opto-mechanics. For example, we
  have transferred the signals from light to mechanical oscillation,
  and vise versa \cite{Verhagen12}. We have realized optomechanically
  induced transparency \cite{Weis10} and generated squeezed lights in
  opto-mechanical systems \cite{Purdy13}.  By using mechanical
  oscillator as an interface, quantum information has been able to
  exchange between microwave superconducting circuits and optical
  lights \cite{Wang12,Andrews14,Yin14}.

  In future we aim to achieve the similar controlling ability of
  macroscopic resonators as we did in the trapped ions. Then we can
  generate Schr\"odinger's cat states with macroscopic objects
  \cite{Romero11}, or even macroscopic living objects, such as virus
  \cite{Romero10}.  As macroscopic quantum superposition states become
  bigger and bigger, we hope to identify the intrinsic decoherence
  mechanisms such as gravity induced decoherence \cite{Penrose96},
  Continuous Spontaneous Localization models
  \cite{Ghirardi86,Ghirardi90}, and et al.. For readers who are
  interested in the theory and developments of optomechanics, please
  read these reviews \cite{optoRev2012,optoRev2013,Liu13}.

  In order to generate the non-classical states in an opto-mechanical
  oscillator, strongly quadratic coupling between the oscillator and
  the optical mode is essential \cite{Romero11}. However, it is
  usually very challenging to get the large quadratic coupling in
  experiments.  Inspired by the trapped ion experiments, the quadratic
  coupling can be replaced by coupling the mechanical oscillator to a
  two-level quantum system, which can be controlled well
  externally. In order to get the stable macroscopic quantum states,
  the coherence time of both the mechanical oscillator and the spin
  $1/2$ should be maximized. There are several proposals that based on
  quantum dots \cite{Wilson04,Bennett10}, atoms \cite{Hammerer09},
  superconducting qubits \cite{OptoGround10} and Nitrogen-vacancy
  centers (NV centers)
  \cite{Rabl09,Rabl10,Arcizet11,Kolkowitz12}. Among all these
  proposals, the ones with NV centers are most attractive. As NV
  centers have presented the coherence even at room temperature.

\begin{figure}[H]
  \centering
  \includegraphics[width=4cm]{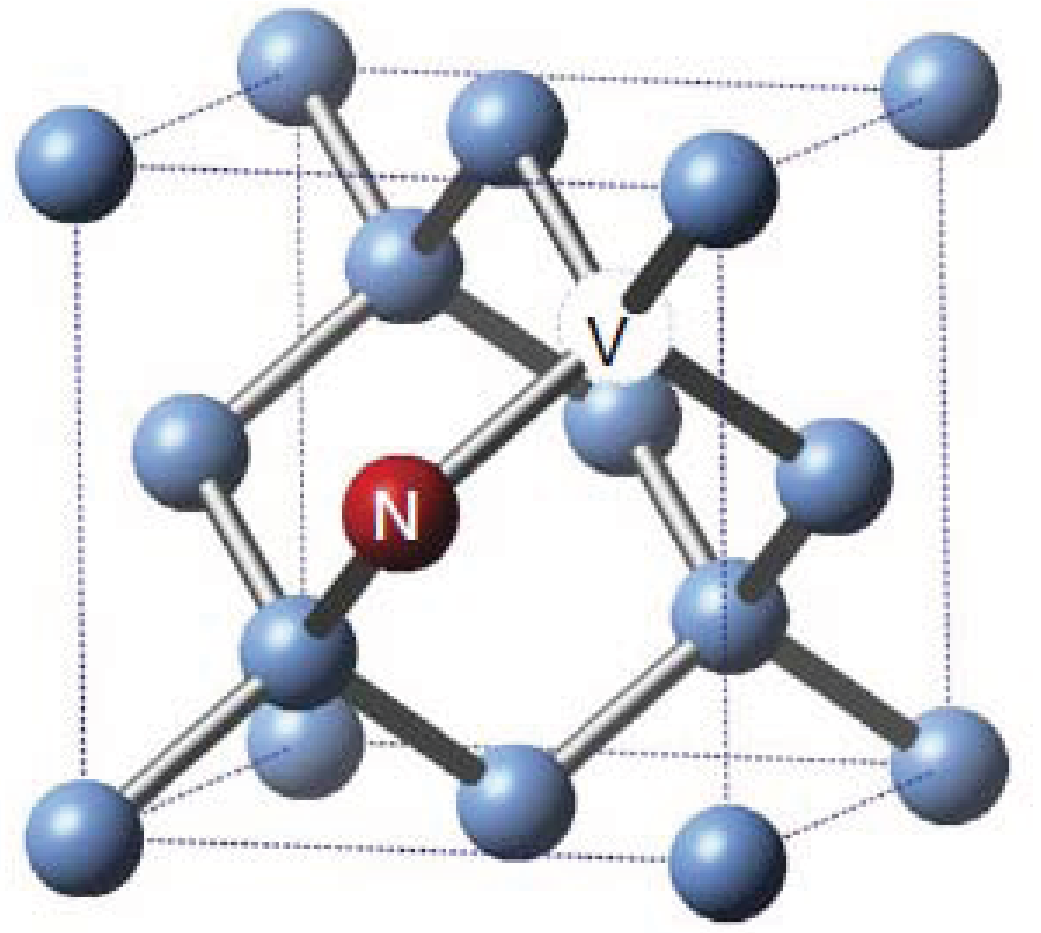}
 \includegraphics[width=4cm]{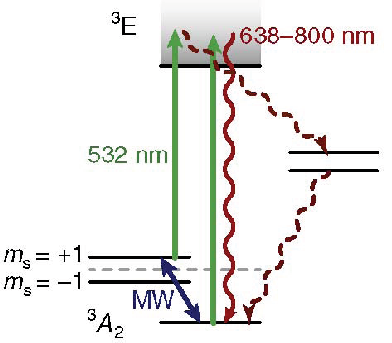}
  \caption{Left figure shows lattice structure of diamond with nitrogen-vacancy center.
  Right one shows the negatively
  charged NV center electronic energy level structure. Electronic spin polarization
  and readout is performed by optical excitation at $532$ nm and red fluorescence
  detection around $637$ nm. Ground-state spin manipulation is achieved by resonant microwave
  excitation. The ground-state triplet has a zero magnetic field splitting $\simeq 2.87$ GHz. Figure
  adopted from Ref. \cite{Bar-Gill12}. Copyright (2012) by Nature Publishing Group.} \label{fig:NVlevel}
\end{figure}

  NV centers in diamond are usually regarded as artificial atoms in
  solid systems.  The diamond lattice consists of covalently bond
  carbon atoms, which makes diamond very stiff.  The valence electrons
  in diamond have huge bandgap ($5.48$ ev), which makes it transparent
  deep into the UV.  In its lattice, nitrogen and vacancy are the most
  common defects. As shown in Fig. \ref{fig:NVlevel},  an NV center
  consists of a nitrogen atom and a
  vacancy at the nearest neighboring site.  The NV centers are usually
  negatively charged, possessing $6$ electrons and spin $S=1$ in the
  ground state.

  NV centers are promising candidate system for quantum information
  processing.  Single NV centers can be addressed using the confocal
  microscopy technique.  The spin state of NV centers can be
  initialized and read-out at room temperature, respectively, by
  optical pumping and spin-dependent fluorescence.  Furthermore, the
  spin state manipulation can be achieved by resonant microwave
  radiation.  Because of the weak spin-orbital coupling of diamond
  material, and low concentration of $^{13}$C, the only spin-carrying
  isotope of carbon with natural abundance $1.1\%$, electron spin of
  NV centers in diamond has very long coherence time ($\sim~\rm{ms}$)
  even at room temperature \cite{Bala09}.  All these properties mentioned above make
  NV centers excellent candidate for quantum logic elements for
  quantum information processing.

  In addition to the applications in quantum information, NV centers
  are widely used as solid-state ultra sensitive magnetic field
  sensor.  Again, because of the long coherence time, a tiny change of
  the magnetic field can be monitored by measuring its effect on the
  spin dynamics of NV centers. Recent experiments demonstrated high
  sensitivity $\sim~\rm{nT}/\sqrt{\rm Hz}$ at atomic scale resolution \cite{Zhao12,Shi14}.
  Furthermore, recent research shows that, not only magnetic fields,
  other types of signal (e.g., electric fields, temperature, and
  strain etc.) which can be converted to magnetic signal can be also
  precisely measured by NV center spins.  This makes NV centers very
  amazing multi-functional atomic-scale sensors.

  There are two methods to interface the NV centers and the mechanical
  oscillators.  The first one is based on strong magnetic field
  gradient to couple the mechanical oscillators with NV centers
  \cite{Rabl09}.  The second one requires the strain induced effective
  electric field to mix phonon mode with NV centers electron spins
  \cite{Bennett13}. The strong coupling may reach in both of the
  methods.  Once the strong coupling regime reaches, we may cooling
  the mechanical oscillator with coherent excitation exchange between
  spin and motion degrees. We can generate arbitrary superposition
  states of mechanical oscillator.  We may realize phonon laser and
  squeezing in the system \cite{Bennett13,Kepesidis13}. The system can
  also be used for quantum information processing \cite{Rabl10}, as it
  can interface many different degree of freedom.

  To increase the coherence time of mechanical oscillator, we can
  optimize the design and choose the materials with high mechanical
  $Q$, such as $\text{Si}_3\text{N}_4$ with $Q$ around $10^6$
  \cite{Zwickl09}.  However, the best way is optically trapping the
  nano-diamond in vacuum \cite{Li11,Kiesel13,Nie13,Nie14,Gieseler12,Yin11,Yin13a,Chang10,Romero10,Neukirch14}.  The
  mechanical $Q$ factor in this system could be as high as $10^{10}$,
  which is comparable to the trapped ions. It is possible to cool the
  trapped nanoparticle from room temperature \cite{Liu14}.
   Another advantage is that
  the strong coupling between the oscillation and the spin can be
  realized with a modest magnetic field gradient around $10^5$ T/m. It
  is also possible to couple the rotation degree of freedom to the
  spins. Besides, the trapping frequency can be easily tuned, even
  completely turned off, which makes the time-of-flight measurement
  possible \cite{Yin13}.

  In an optical trap, the motion of nano-particle can be cooled to mK
  by feedback without cavity \cite{Li11,Gieseler12}. Ground state may
  reach by means of cavity sideband cooling \cite{Yin11,Kiesel13}, or
  by exchanging excitations between phonon and spin degrees
  \cite{Rabl09,Yin13,Neukirch13}.  Even at thermal states, matter wave interference
  could be observed in this systems \cite{Yin13,Scala13,Asadian14}. By
  detecting the coherence of NV center spin, the ultra-sensitive mass
  spectrometer is realizable even at room temperature \cite{Zhao14}.

  This review is organized as follows. In Sec. \ref{sec:Mec}, we will
  review the methods and applications of strong coupling between
  mechanical oscillator and NV centers. In Sec. \ref{sec:Opt}, we will
  review how to observe macroscopic physics and achive ultra-sensitive
  detection in optically trapped nano-diamond that hosts NV centers.
  In Sec. \ref{sec:Con}, we will summarize and discuss the potential
  developments in the field.

  \section{Nanomechanical cantilevers coupled with NV centers} \label{sec:Mec}

  To couple NV center electron spins with mechanical resonator, we
  need to apply additional magnetic field gradient, or induce a strain
  in the lattice of diamond. Here we will briefly discuss these two
  different methods, and illustrate several applications for both of
  the
  methods. 

  \subsection{Magnetic field gradient induced coupling}
  \subsubsection{The setup} \label{sec:rablsetup}
  \begin{figure}[H]
    \centering
    \includegraphics[width=7cm]{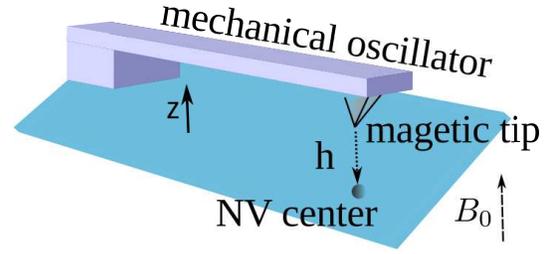}\\
    \caption{A magnetic tip, which is attached to the end of a
      cantilever, is placed at a distance $h$ above a single NV
      center. It creats a strong magnetic gradient near the electronic
      spin of the NV center.  As the spin energy is in proportional to
      the magnetic field, a strong coupling between the NV center
      electrons spin and the motion of cantilever is created. Laser
      field is used for initializing and measuring the spin states.
      Microwave is used for manipulating the the spin
      states.}\label{fig:1}
  \end{figure}

  The idea of strong magnetic coupling between a mechanical oscillator
  and a NV center is shown in Fig. \ref{fig:1} \cite{Rabl09}. Here a
  single NV center electron spin is used for sensing the motion of the
  mechanical oscillator. It should be noted that the NV center can
  also be attached to the end of resonator \cite{Arcizet11}, other
  than the magnetic tip shown in Fig \ref{fig:1}. The frequency of the
  mechanical oscillator is $\omega_r$. The oscillation induces a
  time-varying magnetic field which causes Zeeman shifts of the
  electron spin. The single phonon induces frequency shift $\lambda =
  g_s \mu_b G_m a_0$, where $g_s\simeq 2$, $\mu_B$ is Bohr magneton,
  $G_m$ is the magnetic field gradient, and
  $a_0=\sqrt{\hbar/2m\omega_r}$ is the zero field fluctuation for a
  resonator of mass $m$. We will show that in the current experiments,
  $\lambda$ can be as high as $100$ kHz, which is larger than both the
  spin coherence time ($1$ ms), and the damping rate $\gamma=
  \omega_r/Q$ of high $Q$ mechanical resonators.

  The Hamiltonian of the system can be written as
  \begin{equation}\label{eq:HS}
    H_S= H_{NV} + \hbar \omega_r a^\dagger a + \hbar \lambda (a+ a^\dagger) S_z,
  \end{equation}
  where $H_{NV}$ describes the dynamics of NV center electron spin,
  $a(a^\dagger)$ is the annihilation(creation) operator of the phonon
  in mechanical oscillator, and $S_z$ is the $z$ component of the
  spin-$1$ operator of NV center. We assume that the NV sysmetry axis
  is also aligned along $z$ direction.  The NV center electron spin
  ground state is an $S=1$ triplet.  The three states are labled as
  $|0\rangle, |\pm 1\rangle$. Without external magnetic field, the
  levels $|\pm 1\rangle$ are degenerate. The zero field splitting
  between $|0\rangle$ and $|\pm 1\rangle$ is $\omega_0/2\pi =2.88$
  GHz. In order to manipulate the levels $|\pm 1\rangle$ individually,
  we add a static magnetic field $B_0$ along $z$ axis to break the
  degenerate states $|\pm 1\rangle$. The frequencies difference
  between $|0\rangle$ and $|\pm 1\rangle$ are $\omega_{\pm 1}=\omega_0
  \pm \mu_B B_0/\hbar$.

  In order to coherent exchange the excitation between mechanical
  resonator and the spin, we need to add microwaves to drive Rabi
  oscillations between $|0\rangle$ and $|\pm 1\rangle$. We assume the
  magnetic field oscillation amplitudes is much less than $\hbar
  \omega_0/\mu_B$. In rotating wave frame we have
  \begin{equation}\label{eq:HNV}
    H_{NV}= \sum_{i=\pm 1} -\hbar \Delta_i |i\rangle \langle i | + \frac{\hbar \Omega_i}{2} (|0\rangle \langle i| + |i\rangle \langle 0|),
  \end{equation}
  where $\Delta_{\pm 1}$ and $\Omega_{\pm 1}$ denote the detunings and
  the Rabi frequencies of the two microwave transitions. For
  simplicity, we assume that $\Delta_{+1}=\Delta_{-1}=\Delta$ and
  $\Omega_{+1}=\Omega_{-1}=\Omega$.  The eigenstates of $N_{NV}$ are
  $|d\rangle = (|-1\rangle- |+1\rangle)/\sqrt{2}$, $|g\rangle = \cos
  \theta |0\rangle - \sin \theta |b\rangle$, and $|e\rangle = \cos
  \theta |b\rangle + \sin \theta |0\rangle$, where $|d\rangle
  =(|+1\rangle + |-1\rangle)\sqrt{2}$, and $\tan (2\theta) = -\sqrt{2}
  \Omega/\Delta$. The eigenvalues corresponding to are $\omega_d =
  -\Delta$, and $\omega_{e,g}= (-\Delta \pm \sqrt{\Delta^2 +
    2\Omega^2})/2$. In the case that $\Delta<0$, the lowest energy
  state is $|g\rangle$. We adjust the frequency difference between
  $|d\rangle$ and $|g\rangle$ $\omega_{dg} = \omega_d-\omega_g$ is
  equal to the mechanical frequency $\omega_r$, and the frequency
  difference between $|e\rangle$ and $|d\rangle$ $\omega_{ed}=\omega_e
  - \omega_d$ is largely detuned with $\omega_r$. Under the condition
  $|\omega_r-\omega_{ed}| \gg \lambda$, we will have effective
  Jaynes-Cummings Hamiltonian
  \begin{equation}\label{eq:HJC}
    H_{JC}= \hbar \lambda_g |g\rangle \langle e | a^\dagger + \mathrm{h.c.},
  \end{equation}
  where $\lambda_g = -\lambda \sin \theta$.

  In this system, phonon-spin coupling strength $\lambda$ is usually
  much less than phonon frequency $\omega_r$, which is around $10$
  MHz. Therefore, it is impossible to directly generate
  Schr\"odinger's cat states in this setup. We will discuss how to
  realize frequency tunable mechanical oscillator in the next section.
  By tuning the mechanical frequency less than $\lambda$, the
  Schor\"odinger's cat states in macroscopic mechanical resonator can
  be prepared with spin-dependent force \cite{Yin13}.

  \subsubsection{Applications as resonator
    cooling} \label{sec:cooling} With JC Hamiltonian
  Eq. \eqref{eq:HJC}, we can coherently manipulate the mechanical
  states with the help of spin states if strong coupling condition is
  fulfilled. The decay of mechanical resonator at temperature $T$ is
  defined as $\Gamma_r = k_B T/\hbar Q$. The spin dephasing rate is
  $\delta \omega_{dg}$.  The strong coupling condition is $\lambda_g >
  \Gamma_r, \delta \omega_{dg}$.  In experiments, we may use Si
  cantilever of dimensions $(l,w,t)= (3,0.05,0.05)$ $\mu$m with
  frequency $\omega_r= 7$ MHz and $a_0= 5 \times 10^{-13}$ m. A
  magnetic tip can induce gradient field $G_m = 10^7$ T/m at the
  distant $h=25$ nm. The phonon-spin coupling strength $\lambda = 100$
  kHz is realizable. The mechanical Q can be as high as $10^5$ and the
  heating rate is $\Gamma_r/2\pi =20$ kHz.  The spin decay can be
  around $\delta \omega_{gd} =1$ kHz if we use Carbon-$13$ purified
  diamond. Therefore, the strong coupling condition can be fulfilled.

\begin{figure}[H]
  \centering
  \includegraphics[width=8cm]{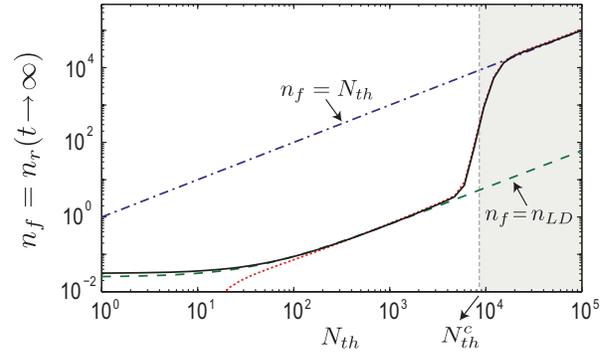}
  \caption{ Steady state phonon occupation number $n_f$ as a function
    of environment thermal excitation $N_{th}$.  The value of
    $N_{th}^c$ denotes the crossover from LD regime to a regime where
    phonon cooling is strongly suppressed and finally $n_f =
    N_{th}$. Firgure adopted from Ref. \cite{Rabl10b}. Copyright
    (2010) by The American Physical Society. }\label{fig:cooling}
\end{figure}

To manipulate the phonon states of the resonator, the first step is to
cool it to the ground state. If the mechanical oscillator frequency is
high enough, e.g. $>1$ GHz, we can use the traditional cryogenic
techniques to cool the environment to $10$ mK, and the resonator is
already in its quantum ground state. Usually the resonator frequency
is around MHz, we have to use active method to cool it to ground
state. In opto-mechanics, we need cavity mode as an vacuum bath to
cool the mechanical mode. In this setup, we don't need cavity, as NV
center can be used as an effective vacuum bath.  By using JC
Hamiltonian \eqref{eq:HJC} and optical pump technology of NV center
spin states, we can cool the resonator to the ground state. The basic
idea is as follows. We initialize the NV center spin to $|0\rangle$ by
optical pump in the time scale less than $1$ $\mu$s. Then we use
microwave to prepare the spin state to $|g\rangle$. By turning on the
microwave driving we use JC Hamiltonian \eqref{eq:HJC} to exchange
transfer excitation from resonator to the NV center spin in the time
scale $1/\lambda_g \sim 10~\mu$s. As last we initialize the NV center
again. By repeating the process, the excitations in resonator are
quickly removed. Under the condition $\lambda_g > \Gamma_r$, the
quantum ground state cooling should be possible. If we carefully
adjust the detuning of driving microwave and laser, the above
procedure can automatically repeated to cool the mechanical resonator
\cite{Rabl09}.

The above cooling method is only valid for low temperature $T<1$ K and
mechanical frequency $\omega_r \sim 1$ MHz. For high temperature, the
Lamb-Dicke(LD) limit is not fulfilled, the cooling with NV center
spins is no longer possible \cite{Rabl10b}, as shown in Fig.
\ref{fig:cooling}. When thermal phonon phonon is larger than $N_{th}^c
=10^4$, the cooling effects quickly disappear.  After quantum ground
state being prepared, arbitrary superposition states of resonator
should be possible. However, to generate the high fidelity phonon
states, we need a high $Q$ resonator, which is very difficult for this
system. We will discuss this problem in Sec. \ref{sec:Opt}.

\subsubsection{Applications in quantum information
  processing} \label{sec:QI}
Once mechanical ground state is prepared, the mechanical resonator can
be used for quantum information processing
\cite{Xu10,Chen10,Zhou10}. By charging the mechanical oscillator,
distant coupling between resonators is also possible \cite{Rabl10}.

\begin{figure}[H]
  \centering
  \includegraphics[width=8cm]{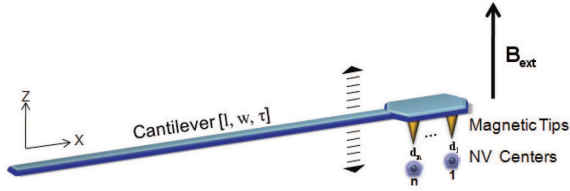}\\
  \caption{ A scheme to entangle multiple NV centers spins that
    couple with the same mechanical oscillator.  (a) An array of
    evenly spaced magnetic tips are attached at the end of a
    nanomechanical cantilever, under which are one-to-one
    correspondent NV centers. Figure adapted from
    Ref. \cite{Xu10}.
Copyright (2010) by The American
    Physical Society.}\label{fig:xu}
\end{figure}

In order to entangle distant NV centers, here we need to couple
multiple NV centers to the same mechanical oscillator, which is used
as a quantum bus. As shown in Fig. \ref{fig:xu}a, Xu and et
al. considered the scheme that an array of a NV centers locate equally
distance, above which are one-to-one magnetic tips attached at the end
a cantilever. The qubits are encoded in two spins levels $0\rangle$
and $-1\rangle$.  We applied a microwave to globally drive all the NV
center spins. We assume that all NV center spins have the same
transition frequency $\omega_0$ and the frequency of microwave is on
resonant with $\omega_0$. We define the new basis for the spins as
$|\pm \rangle = (|-1\rangle + |+1\rangle)/\sqrt{2}$. Consider the case
that the Rabi frequency of microwave $\Omega$ is comparable to the
mechanical frequency $\omega_r$, in rotating wave approximation, we
have the effective Hamiltonian of the system as
\begin{equation}\label{eq:Xu}
  H_I = \sum_{j=1}^n \frac{\lambda}{2} a|+\rangle_j \langle -| \exp [-i\delta (t) dt] + \mathrm{h.c.},
\end{equation}
with $\delta (t) = \omega_r - \Omega(t) $. The collective Hamiltonian
Eq. \eqref{eq:Xu} can be used for generating W states. As an example,
initially we prepare the phonon state to the Fock state $|1\rangle$
with phonon number $1$, and the NV centers to the state
$|-\rangle_1|-\rangle_2 \cdots |-\rangle_n$. We set the detuning
$\delta(t) = 0$, the W state $|W\rangle = (|+\rangle_1|-\rangle_2
\cdots |-\rangle_n+|-\rangle_1|+\rangle_2 \cdots |-\rangle_n \cdots
|-\rangle_1|-\rangle_2 \cdots |+\rangle_n )/\sqrt{n}$ can be generated
at the time $t= \pi/(\sqrt{n} \lambda)$.

If the detuning $\delta(t)$ is much larger than coupling strength
$\lambda$, it will induce the effective spin-spin coupling between NV
centers \cite{Zheng00,Zhou10}. In Ref. \cite{Zhou10}, they only
considered two NV centers coupling with the same cantilever. Assume the
detuning $\delta_1= \delta_2=\delta$, to the second order we have the
effective Hamiltonian
\begin{equation}\label{eq:Zhou}
  H_{eff}= \frac{\hbar \lambda_g^2}{4\delta} [(a^\dagger a +\frac{1}{2}) (\sigma_{1z} + \sigma_{2z}) - (\sigma_{1+}
  \sigma_{2-} + \sigma_{1-}\sigma_{2+})].
\end{equation}
As we discussed in Sec. \ref{sec:cooling}, the phonon-spin coupling
$\hbar \lambda$ could be around $100$ kHz. In order to fulfill the
condition $\delta \gg \lambda_g$, we choose $\hbar \delta$ in the
order of $1$ MHz. The spin-spin coupling $\hbar \lambda_g^2/(4\delta)$
is in the order of $10$ kHz, which is larger than both NV center spin
decoherence rate and the effective phonon decay rate at $10$ mK
temperature. Therefore, it is possible to generate the entanglement
among NV centers spins in experiments with high fidelity.

\begin{figure}[H]
  \centering
  \includegraphics[width=8cm]{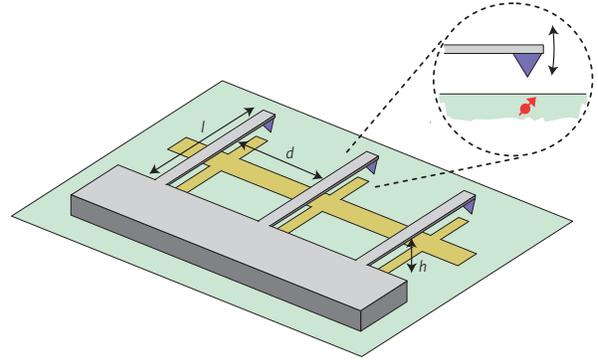}
  \caption{ Schematic view of a scalable quantum information processor
    on an electromechanical quantum bus and NV centers spins as
    qubits. Figure adapted from Ref. \cite{Rabl10}. Copyright (2010) by Nature Publishing Group.
  }\label{fig:transducer}
\end{figure}

The idea can be extended to realize a scalable quantum information
processor \cite{Rabl10}.  As shown in Fig. \ref{fig:transducer}, the
processor consist an array of $N$ nanomechanical resonators, each of
which is coupled magnetically to an electronic spin qubit associated
with an impurity located in the substrate below. As we discussed in
Sec. \ref{sec:rablsetup}, the resulting spin-phonon coupling could be
as high as $50$ kHz under the magnetic gradient $10^7$ T/m.  The long
range interaction between distant sites can be established by charging
the resonator and capacitively coupling with the nearby wire. The
Hamiltonian of the resonators is
\begin{equation}\label{eq:coupled}
  H_{cr} = \sum_i \hbar \omega_r a^\dagger_i a_i + \frac{\hbar}{2} \sum_{i,j} g_{ij} (a_i + a_i^\dagger)
  (a_j +a_j^\dagger)= \sum_n \hbar\omega_n a_n^\dagger a_n
\end{equation}
After combining magnetic and electric coupling Hamiltonians, we have
the full Hamiltonian of the system,
\begin{equation}\label{eq:totalcoupled}
  H_{cr} =H_s(t) + \hbar \sum_n \hbar\omega_n a_n^\dagger a_n +\frac{1}{2} \sum_{i,n} \lambda_{i,n} (a_n^\dagger + a_n) \sigma_z^i,
\end{equation}
where $H_st) = \sum_i \hbar (\delta_i \sigma_z^i + \Omega_i(t)
\sigma_x^i)/2$, $\omega_n$ and $a_n$ denote frequencies and collective
mode operators for phonon eigenmodes. This model is quite similar as the quantum computation
proposal for trapped ions systems. However, here the decoherence
mechanism and physical implementation are different. Anyway, the idea
of using hybrid systems that contain mechanical resonators array and
NV centers for quantum information processing needs to be further
investigated in future.

\subsubsection{Applications in ultra-sensitive measurement}
As the resonator is coupled with NV center spin, we can use NV center
as a detector to measure the state of resonator
\cite{Arcizet11,Kolkowitz12}. Therefore, the NV center spin can be
used as an ultra-sensitive detector for mechanical resonators. Even
strong coupling conditions is not fulfilled, we may still detect the
Brownian motion of resonator in classical regime.

\begin{figure}[H]
  \centering
  \includegraphics[width=8cm]{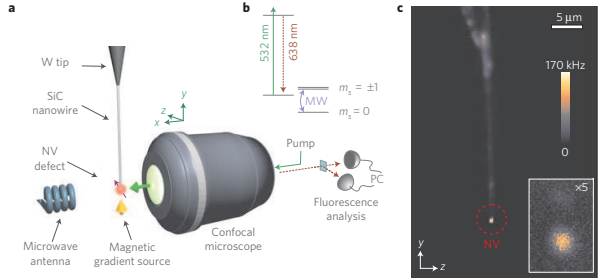}\\
  \caption{(a) A confocal microscope monitors the fluorescence of
  a single NV defect hosted in a diamond nanocrystal placed at the
extremity of a SiC nanowire. A microwave antenna is used to
manipulate the NV electronic spin, while a micro-fabricated magnetic structure
approached in the vicinity of the suspended NV centre generates a
strong magnetic field gradient. (b), The  electronic levels structure
of the NV centres at zero magnetic field. (c), Fluorescence map of the system recorded with
the confocal microscope while scanning the objective position. The isolated bright
spot circled in red corresponds to the fluorescence of a single NV
centre. Inset: zoom on the nanowire extremity. Figure adapted from \cite{Arcizet11}.
Copyright (2011) by Nature Publishing Group.
}\label{fig:nphysics}
\end{figure}
In Ref. \cite{Arcizet11}, it realized the hybrid system that
a single nitrogen-vacancy center coupled with a nanomechanical
oscillator. A single NV center hosted in a nano-diamond is placed
at the extremity of a Sic nanowire. By adding a strong magnetic field
gradient to the system, a magnetic coupling between NV center and
mechanical oscillator is induced. The motion of nanowire can be
probed by reading out a single electron spin of the NV center.

Reference \cite{Kolkowitz12} reported the coherent coupling of a mechanical
cantilever to the single spin of NV center. The authors used
the electronic spin of an NV center to sense the motion of
a magnetized cantilever. Although the magnetic field change
due to the motion of the cantilever is very small, it can still be
monitored by measuring the NV center spin coherence. This
was achieve with the help of the spin echo technique, where
the spin is flipped by microwave $\pi$ pulses during the evolution.
With this technique, the unwanted background noise
is filtered out, while the signal to be detected (i.e., the magnetic
field signal from the cantilever motion) is magnified.
In Ref. \cite{Kolkowitz12}, the authors successfully demonstrated a motion
sensing with precision down to $\sim$ pm under ambient conditions.
Later, we will show the spin-echo-based technique is
not limit in motion sensing. Indeed, it can be applied in sensing
various types of signals \cite{Zhao14}.

\subsection{Strain induced coupling}

\subsubsection{Model}

In the previous subsection, we summarized the recent works on hybrid
systems with a mechanical resonator and a NV center, where a strong
magnetic gradient induces spin-phonon coupling.  In this section, we
will discuss new a mechanism that couples NV center electrons and the
the phonon mode in diamond, strain induced spin-phonon coupling, and
effective spin-spin interactions between NV centers
\cite{Bennett13,Teissier14}. The spin-spin interactions can induce
spin squeezing in NV center ensembles (NVE), which may be used in NVE
magnetometry.

\begin{figure}[H]
  \centering
  \includegraphics[width=8cm]{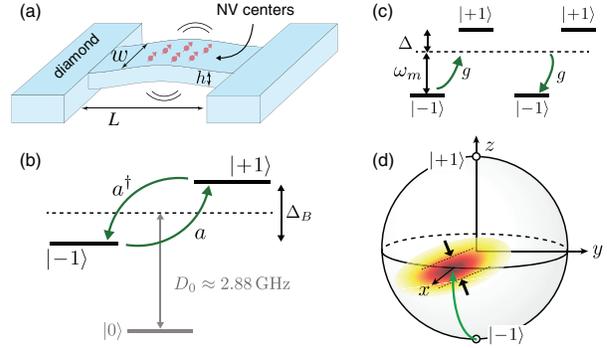}
  \caption{ (a) All-diamond doubly clamped mechanical resonator with
    an embedded NV centers ensemble (NVE). (b) Spin levels and
    transitions of NV centers electron spins. Local perpendicular
    strain induced by beam bending mixes the $|\pm 1\rangle$
    states. (c) NV centers spins in the two-level subspace $|\pm
    1\rangle$ off-resonantly couple with a common mechanical mode. (d)
    NVE collective spins squeezing.  Figure adopted from
    Ref. \cite{Bennett13}. Copyright (2013) by The American Physical
    Society.}\label{fig:strain}
\end{figure}

As shown in Fig. \ref{fig:strain}(a), we consider an NVE embedded in a
single crystal diamond nanobeam.  When beam oscillates, it strains the
lattice of diamond, and induces direct coupling between NV center
electrons spins and the mechanical mode.  Here we write down the
Hamiltonian for NV center, in the presence of external electric and
magnetic fields $\vec{E}$ and $\vec{B}$ \cite{Doherty12,Bennett13}
\begin{equation}\label{eq:strain}
  \begin{aligned}
    H_{\mathrm{NV}}= &(D_0 + d_{\|}E_z) S_z^2 + \mu_B g_s \vec{S} \cdot \vec{B} \\
    &- d_{\perp} [E_x(S_x S_y +S_yS_x) + E_x (S_x^2 - S_y^2)],
  \end{aligned}
\end{equation}
where $D_0/2\pi = 2.88$ GHz is the zero field splitting, $g_s\simeq
2$, $\mu_B$ is the Bohr magneton, and $d_{\|}$ ($d_{\perp}$) is the
ground state electric dipole moment in the direction parallel
(Perpendicular) to the NV axis \cite{Dolde11}.

As shown in Fig. \ref{fig:strain}(b), motion of nanoresonator changes
the local strain of the NV center, and induces an effective electric
field. We are interested to the near resonant coupling between a
single motion mode and $|\pm 1\rangle$ transition of the NV
center. The Zeeman splitting between $|\pm 1\rangle$ is $\Delta_B =
g_s \mu_B B_z/ \hbar$. The perpendicular component of strain
$E_{\perp}$ mixes $|\pm 1\rangle$ states with interaction Hamiltonian
$E_{\perp}= E_0 (a+ a^\dagger)$, where $a$ is the destruction operator
of the resonator mode with frequency $\omega_m$.  $E_0$ is the zero
point motion of the beam induced perpendicular strain. The parallel
strain also induces the level shifts between $|\pm 1\rangle$ and
$|0\rangle$. However, in the subspace $|\pm 1\rangle$ the parallel
strain plays no role in the system, as state $|0r\rangle$ is not
involved. In this subspace, the interaction Hamiltonian of each NV
center is
\begin{equation}\label{eq:strainIn}
  H_I =\hbar g(a^\dagger \sigma_i^- + a\sigma_i^+),
\end{equation}
where $\sigma_i^\pm = |\pm 1\rangle \langle \mp 1|$ is Pauli operator
for $i$th NV center and $g$ is the single phonon coupling
strength. For NVE, we can define the collective spin operators, $J_z =
\frac{1}{2} \sum_i |1\rangle_i \langle 1| - |-1\rangle_i \langle -1 |$
and $J_\pm = J_x \pm iJ_y= \sum_i \sigma_i^\pm$. The total Hamiltonian
is
\begin{equation}\label{eq:straintotal}
  H_T= \omega_m a^\dagger a + \Delta_B J_z + g( a^\dagger J_- + a J_+).
\end{equation}
Here for simplicity, we assume that coupling $g$ is uniform for all NV
centers.  The interaction between NV centers is also neglected as we
assume that they are far apart from each other.

The coupling strength $g$ can be estimated with the follow equation
\cite{Bennett13}
\begin{equation}\label{eq:straing}
  \frac{g}{2\pi} \approx 180 \big( \frac{\hbar}{L^3 w \sqrt{\rho E}} \big)^{1/2},
\end{equation}
where $\rho$ is the mass density and $E$ is the Young's modulus of
diamond.  Up to now, we only achieved the single phonon coupling
strength $g \sim 0.25$ Hz in a diamond resonator \cite{Teissier14}.
Theoretically, we may get $g\sim 1$ kHz in future.  For a beam of
dimensions $(L,w,h)=(1,0.1,0.1)~\mu$m we got $\omega_m/2\pi \sim 1$
GHz and coupling $g/2\pi \sim 1$ kHz. We define the single spin
cooperativity $\eta = g^2 T_2/(\gamma \bar{n}_{th})$, where
$\gamma=\omega_m/Q$ is the mechanical decay rate and $T_2$ is the spin
dephasing time, and $\bar{n}_{th}$ is the thermal equilibrium phonon
number for resonator. It is found that $\eta$ could be as high as
$0.8$ for $T=4$ K, $T_2=10$ ms, and $Q=10^6$. To get $\eta>1$, we need
to further decrease the dimension of the diamond resonator, or
decrease the temperature $T$.  Similarly as Sec., the Hamiltonian can
also be used for cooling the mechanical oscillator if the strong
coupling condition is fulfilled \cite{Kepesidis13}.

\subsubsection{Spin squeezing }

Similarly as we discussed in Sec. \ref{sec:QI} \cite{Zhou10,Zheng00},
in large detuning limit that $g \ll \Delta = \Delta_B - \omega_m$, we
can adiabatically eliminate the mechanical mode $a$, and get effective
spin-spin interactions \cite{Bennett13}.  The effective Hamiltonian is
\begin{equation}\label{eq:squeezing}
  H_{eff}= \omega_m a^\dagger a + (\Delta_B + \lambda a^\dagger )J_z + \frac{\lambda}{2} J_+J_-,
\end{equation}
where $\lambda = 2g^2/\Delta$. The Hamiltonian
Eq. \eqref{eq:squeezing} can be used for generating spin squeezing
state. To generate the spin squeezed state, we initialized the NVE to
coherent spin state $|\psi_0\rangle$ along $x$ axis which satisfies
$J_x |\psi_0 \rangle = J |\psi_0\rangle$. It has equal transverse
variances $\langle J_x^2 \rangle = \langle J_x^2 \rangle = J/2$.  The
interaction term $J_+ J_-= \textbf{J}^2 - J_x^2 + J_z$, where total
angular momentum $J$ is conserved and $J_z^2$ induces spin variance
squeezing in one direction, as shown in Fig. \ref{fig:strain}(d). The
squeezing can be qualified by the parameter
\begin{equation}\label{eq:squeezeP}
  \xi^2 = \frac{2J\langle \Delta J^2_{\text{min}}\rangle}{\langle J_x \rangle^2} ,
\end{equation}
where $\langle \Delta J_{\text{min}}^2 \rangle = \frac{1}{2} (V_+ -
\sqrt{v_-^2 + V_{yz}^2})$ is the minimum spin uncertainty with $V_\pm
= \langle J_y^2 \pm J_z^2 \rangle$ and $V_{yz} = \langle J_y J_z + J_z
J_y \rangle /2$. If we prepare the spin squeezed state with $\xi^2
<1$, it has a direct applications for magnetometry of NVE below the
projection noise ilmit \cite{Ma11}.

\begin{figure}[H]
  \centering
  \includegraphics[width=8cm]{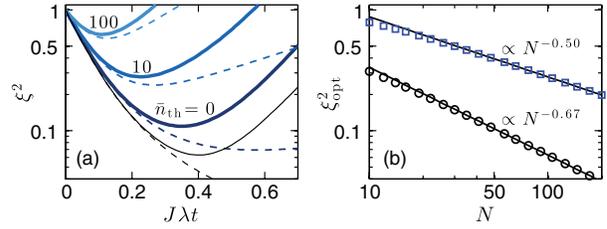}
  \caption{ (a) Spin squeezing parameter versus scaled precession time
    with $N=100$ spins. Thick blue (gray) lines show the calculated
    squeezing parameter for $T_2=10$ ms and values of
    $\bar{n}_{\mathrm{th}}$ as shown. For each curve, the squeezing is
    optimized. Dashed lines are calculated from the linearized
    equations for the spin operator averages. (b) Optimal squeezing
    versus number of spins. Lower (upper) line shows power law fit for
    $\bar{n}_{\text{th}}= 1(10)$ and $T_2= 1(0.01)$ s . Other
    parameters in both plots are $\omega_m=1$ GHz, $g/2\pi=1$ kHz,
    $Q=10^6$. Figure adopted from Ref. \cite{Bennett13}. Copyright
    (2013) by The American Physical Society.  }\label{fig:squeeze}
\end{figure}

In Fig \ref{fig:squeeze}(a), the squeezing parameter verse time is
plotted for an ensemble of $N=100$ spins and several
$\bar{n}_{\text{th}}$.  The decoherence of spin and mechanical decay
is also considered by solving master equation \cite{Bennett13}. We
plot the scaling of the squeezing parameter with $J$ for small
decoherence in Fig. \ref{fig:squeeze}(b). It was found the scalling is
$\xi^2_{\text{opt}} \sim J^{-2/3}$.

\section{Optically trapped nano-diamond that hosts NV
  centers} \label{sec:Opt} As we stated in the introduction, one of
the key factors affecting macroscopic Schr\"odinger's cat state is the
mechanical Q factor. In this section we will focus on the optically
trapped nanodiamond system, whose mechanical Q factor is not related
the the material, but the pressure of the vacuum. For the current
technology, we can have Q has high as $10^{10}$.  Therefore, the life
time of Schr\"odinger's cat state can be as long as millisecond.

\subsection{Scheme and Fock States Preparation}

\begin{figure}[H]
  \centering
  \includegraphics[width=6cm]{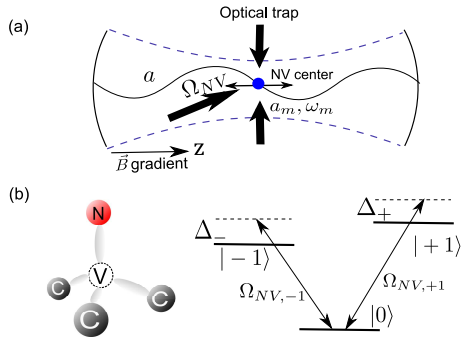}
  \caption{ (a) A nanodiamond with a NV center is optically trapped in
    vacuum with spin-mechanical coupling enabled through a nearby
    magnetic tip and opto-mechanical coupling through a cavity
    around. (b) The atomic structure (left) and the level diagram
    (right) in the ground state manifold for a NV center in the
    nanodiamond. Figure adopted from \cite{Yin13}.  Copyright (2013)
    by The American Physical Society. }\label{fig:yin}
\end{figure}
As shown in Fig. \ref{fig:yin}a, we consider a nano-diamond trapped by
an optical tweezers in a high-Q cavity in vacuum.  Therefore, the
mechanical motion of the nanodiamond couples with the cavity mode. The
trap is located in a place that the cavity mode has the maximum
gradient. Near the NV center, there is a magnet tip, which induces a
strong magnetic field gradient. The magnetic gradient couples the
mechanical motion and the electron spin. There is also a microwave
source to control the spin of the NV center inside the nanodiamond.
As the nanodiamond is optically trapped in the vacuum, the coherence
time of the mechanical mode for its center-of-mass motion is long
\cite{Yin13}. The frequency of the optical trap can be quickly tuned
through control of the laser intensity. This feature is important as
we can cool the mechanical mode to the ground state in a strong trap
and prepare large quantum superposition states of the nanodiamond in a
weak trap by a quench of the trap frequency.

Once the resonator cooled down, we can generate many non-classical
states, such as Fock states. The NV spin is initially set to the state
$|0\rangle $, which is decoupled from the mechanical mode during the
cooling. Initialization and single shot detection of the NV spin have
been well accomplished experimentally \cite{Robledo11}. We assume that
the NV\ center is at a position with zero magnetic field and a large
field gradient. We apply a microwave drive with the Hamiltonian
$H_{drive}=\hbar (\Omega _{\mathrm{NV}%
  ,+1}e^{i\omega _{l+}t}|0\rangle \langle +1|+\Omega _{\mathrm{NV}%
  ,-1}e^{i\omega _{l-}t}|0\rangle \langle -1|+h.c.)/2$ and set the
Rabi frequency $\Omega _{\mathrm{NV},\pm 1}=\Omega _{\mathrm{NV}}$ and
the detuning $\Delta _{\pm }\equiv \omega _{l\pm }-\omega _{\pm
  1}=\Delta $.  With $\Delta \gg |\Omega _{\mathrm{NV}}|$, we
adiabatically eliminate the level $|0\rangle $ and get the following
effective Hamiltonian
\begin{equation}
  \begin{aligned} H_{e} =\hbar \omega_m a_m^\dagger a_m +\hbar\Omega\sigma_z +
    \hbar \lambda (\sigma_+ + \sigma_-) (a_m +a_m^\dagger), \end{aligned}
  \label{eq:Heff1}
\end{equation}%
where $\Omega =|\Omega _{\mathrm{NV}}|^{2}/4\Delta $, $\sigma
_{z}=|+\rangle \langle +|-|-\rangle \langle -|$, $\sigma
_{+}=|+\rangle \langle -|$, $%
\sigma _{-}=|-\rangle \langle +|$, and we have defined the new basis
states $%
|+\rangle =(|+1\rangle +|-1\rangle )/\sqrt{2}$, $|-\rangle
=(|+1\rangle -|-1\rangle )/\sqrt{2}$.  In the limit $\lambda \ll
\omega _{m}$, we set $%
\Omega =\omega _{m}/2$ and use the rotating wave approximation to get
an effective interaction Hamiltonian between the mechanical mode and
the NV center spin, with the form
$$H_{JC}=\hbar \lambda \sigma _{+}a_{m}+h.c..$$
This represents the standard Jaynes-Cummings(J-C) coupling
Hamiltonian.  Similarly, if we set $\Omega =-\omega _{m}/2$, the anti
J-C Hamiltonian can be realized with
$$H_{aJC}=\hbar \lambda \sigma _{+}a_{m}^{\dagger }+h.c..$$

Arbitrary Fock states and their superpositions can be prepared with a
combination of J-C and anti J-C coupling Hamiltonians. For example, to
generate the Fock state $|2\rangle _{m}$, we initialize the state to
$%
|+\rangle |0\rangle _{m}$, turn on the J-C coupling for a duration $%
t_{1}=\pi /(2\lambda )$ to get $|-\rangle |1\rangle _{m}$, and then
turn on the anti J-C coupling for a duration $t_{2}=t_{1}/\sqrt{2}$ to
get $%
|+\rangle |2\rangle _{m}$. The Fock state with arbitrary phonon number
$%
n_{m} $ can be generated by repeating the above two basic steps, and
the interaction time is $t_{i}=t_{1}/\sqrt{i}$ for the $i$th step.
Superpositions of different Fock states can also be generated.  For
instance, if we initialize the state to $(c_{0}|+\rangle
+c_{1}|-\rangle )\otimes |0\rangle _{m}/\sqrt{2}$ through a microwave
with arbitrary coefficients $c_{0},c_{1}$, and turn on the J-C
coupling for a duration $t_{1}$, we get the superposition state
$|-\rangle \otimes (c_{1}|0\rangle _{m}+ic_{0}|1\rangle
_{m})/\sqrt{2}$. Using the optical cavity, the Fock state
$|n_{m}\rangle _{m}$ of mechanical mode can also be mapped to the
corresponding Fock state of the output light field \cite{Yin11}.

The effective QND Hamiltonian for the spin-phonon coupling takes the
form
$$H_{QND}=\hbar \chi \sigma _{z}a_{m}^{\dagger }a_{m},$$
with $\chi =4\Omega \lambda ^{2}/(4\Omega ^{2}-\omega _{m}^{2})$ when
the detuning $||\Omega |-\omega _{m}/2|\gg \lambda $. The Hamiltonian
$H_{QND}$ can be used for a quantum non-demolition measurement(QND)
measurement of the phonon number: we prepare the NV center spin in a
superposition state $|+\rangle +e^{i\phi }|-\rangle )/\sqrt{2}$, and
the phase $\phi $ evolves by $\phi (t)=\phi _{0}+2\chi n_{m}t$, where
$n_{m}=a_{m}^{\dagger }a_{m}$ denotes the phonon number. Through a
measurement of the phase change, one can detect the phonon number.

Let us estimate the typical parameters. A large magnetic field
gradient can be generated by moving the nanodiamond close to a
magnetic tip.  Though magnetic gradient up to $10^7$ T/m has been
realized \cite{Tsang06,Mamin07}, here we take the gradient
$G=10^{5}$~T/m, which is much less than the one used in
Sec. \ref{sec:cooling} \cite{Rabl09}. We get the coupling $%
\lambda \simeq 2\pi \times 52$~kHz for a nanodiamond with the diameter
$d=30$%
~nm in an optical trap with a trapping frequency $\omega _{m}=2\pi
\times 0.5 $~MHz.
The QND detection rate $2|\chi |\sim 2\pi \times 25$~kHz with the
detuning $||\Omega |-\omega _{m}/2|\sim 5\lambda $. The NV electron
spin dephasing time over $1.8$ ms has been observed at room
temperature \cite{Bala09}, which is long compared with the Fock state
preparation time $1/\lambda $ and the detection time $%
1/\left( 2|\chi |\right) $.

\subsection{Macrocopic quantum superpositions}
To prepare spatial quantum superposition state, we need to generate
quantum superposition of the nanodiamond at distinct locations.  To
detect the superposition state, we need to do interference experiment,
either with matter wave \cite{Yin13} or with the spin \cite{Scala13}.
\subsubsection{Matter wave interference}
Without the microwave driving, the spin-mechanical coupling
Hamiltonian takes the form
\begin{equation} \ H=\hbar \omega _{m}a_{m}^{\dagger }a_{m}+\hbar
  \lambda S_{z}(a_{m}+a_{m}^{\dagger }).  \label{eq:Hlow}
\end{equation}%
The mechanical mode is initialized to the vacuum state $|0\rangle_m $
(or a Fock state $|n_{m}\rangle_m $) in a strong trap with the
trapping frequency $%
\omega _{m0}$ and the NV center spin is prepared in the state
$|0\rangle $.  Although the ground state cooling is most effective in
a strong trap, to generate large spatial separation of the wave
packets it is better to first lower the trap frequency by tuning the
laser intensity for the optical trap.  While it is possible to lower
the trap frequency through an adiabatic sweep to keep the phonon state
unchanged, a more effective way is to use a non-adiabatic
state-preserving sweep \cite{Chen10a}, which allows arbitrarily short
sweeping time.  We denote $|n_{m}\rangle _{m1}$ as the mechanical
state in the lower frequency $\omega _{m1}$. We then apply an
impulsive microwave pulse to suddenly change the NV spin to the state
$(|+1\rangle +|-1\rangle )/\sqrt{2}$ and simultaneously decrease the
trap frequency to $\omega _{m2}\leq \omega _{m1}$. The evolution of
the system state under the Hamiltonian \eqref{eq:Hlow} then
automatically split the wave packet for the center-of-mass motion of
the nanodiamond. The splitting attains the maximum at time
$T_{2}/2=\pi /\omega _{m2}$, where the maximum distance of the two
wave packets in the superposition state is $D_{m}=8\lambda
a_{2}/\omega _{m2}=4g_{s}\mu _{B}G/(m\omega _{m2}^{2})$, where
$a_{2}=\sqrt{%
  \hbar /2m\omega _{m2}}$. At this moment, the system state is
\begin{equation}\label{eq:displacement}
  |\Psi_{S}\rangle =(|+1\rangle |D_{m}/2\rangle _{n_{m}}+|-1\rangle
  |-D_{m}/2\rangle _{n_{m}})/\sqrt{2},
\end{equation}
where $|\pm D_{m}/2\rangle _{n_{m}}\equiv (-1)^{a_{m}^{\dagger }a_{m}}
e^{ \pm D_{m}(a_{m}^{\dagger }-a_{m})/4a_{2}} \left\vert
  n_{m}\right\rangle _{1}$ is the displaced Fock state (or coherent
states when $n_{m}=0$).  This is just the entangled spatial
superposition state.

\begin{figure}[H]
  \centering
  \includegraphics[width=4.4cm]{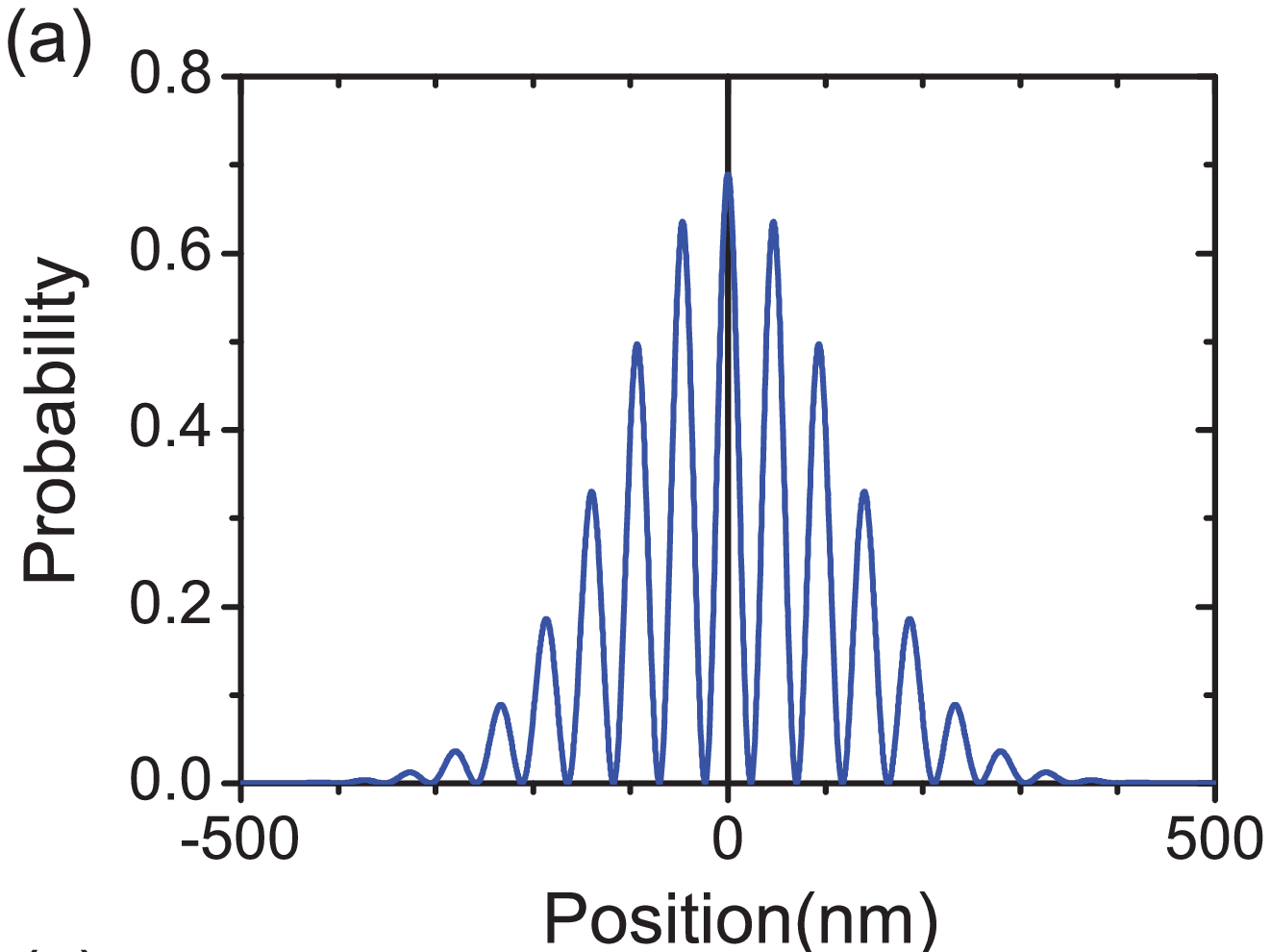}
  \includegraphics[width=3.7cm]{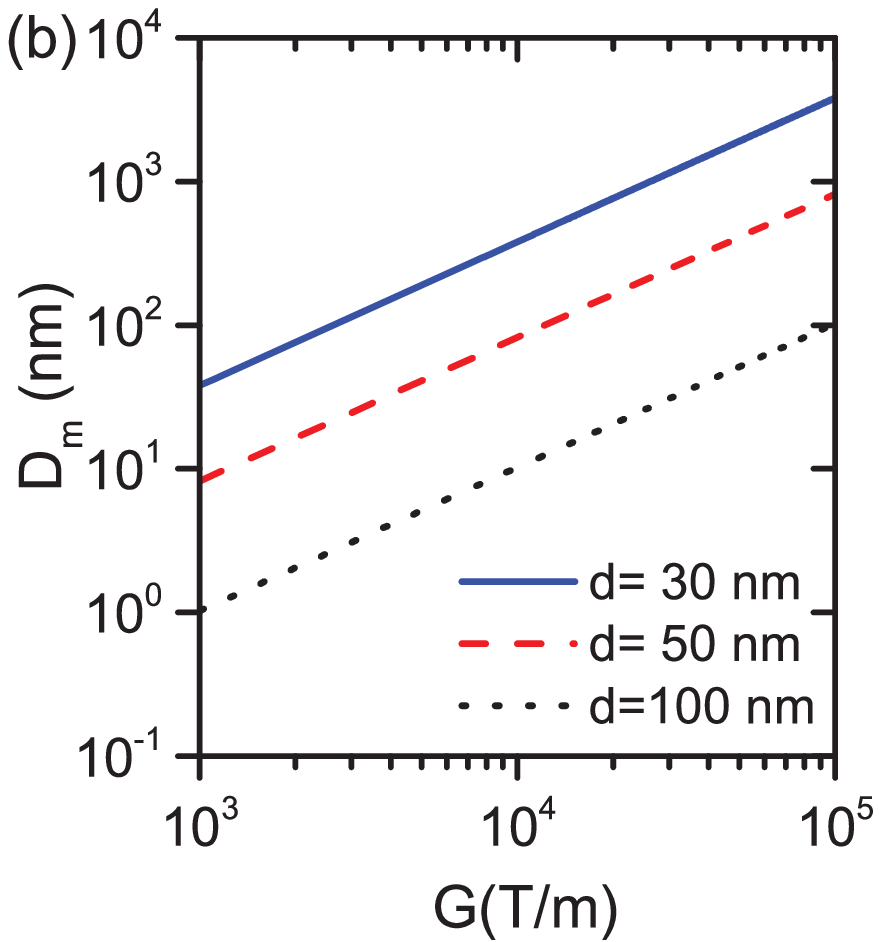}
  \caption{ (a) Spatial interference patterns for a 30~nm nano-diamond
    after 10~ms of free expansion. The nano-diamond is initially
    prepared in the vacuum state $|0 \rangle_{m}$ or the 1-phonon
    state $|1 \rangle_{m}$ of a 20 kHz trap. The magnetic gradient is
    $3 \times 10^4$T/m. Before the trap is turned off, the center of
    mass of the nano-diamond is prepared in $|\protect\psi_+\rangle_0$.
    (b) Maximum spatial separation $D_{m}$ as a function
    of the magnetic gradient $G$ when the trapping frequency is $1$ kHz.
    Macroscopic superposition states with separation larger than the
    size of the particle can be achieved with a moderate magnetic
    gradient. Figure adopted from \cite{Yin13}.  Copyright (2013) by
    The American Physical Society.}\label{fig:yin1}
\end{figure}

The maximum distance $D_{m}$ is plotted in Fig. \ref{fig:yin1}a versus
trap frequency, magnetic field gradient, and diameter $d$ of the
nanodiamond, and superposition states with separation $D_{m}$
comparable to or larger than the diameter $d$ is achievable under
realistic experimental conditions.

To transform the entangled cat state $|\Psi _{S}\rangle $ to the
standard cat state $\left\vert \psi _{\pm }\right\rangle
_{n_{m}}\equiv (|D_{m}/2\rangle _{n_{m}}\pm |-D_{m}/2\rangle
_{n_{m}})/\sqrt{2}$, we need to apply a disentangling operation to
conditionally flip the NV spin using displacement of the diamond as
the control qubit. This can be achieved as different displacements of
the wavepacket induce relative energy shifts of the spin levels due to
the applied magnetic field gradient \cite{Yin13}.  To detect spatial
superposition state, we can turn off the optical trap and let the
spatial wave function freely evolve for some time $t$. The split wave
packets will interference just like the Young's double slit
experiment.  The period of the interference pattern is $\Delta z=2\pi
\hbar t/(mD_{m})$.  As an estimation of typical parameters, we take $%
\omega _{m1}=\omega _{m2}=2\pi \times 20$ kHz, $d=30$ nm, and magnetic
field gradient $3\times 10^{4}$ T/m. The spin-phonon coupling rate
$\lambda \simeq 2\pi \times 77$ kHz and the maximum distance
$D_{m}\simeq 31 a_2$. The preparing time of superposition state is
about $25$ $\mu $s, which is much less than the coherence time of the
NV spin. For the time of flight measurement after turn-off of the
trap, we see the interference pattern with a period of $47$ nm after
$t=10$~ms, as shown in Fig. \ref{fig:yin1}, which is large enough to
be spatially resolved \cite{Li11,Gieseler12}.

\subsubsection{Ramsey interference}

The challenging of matter wave interference proposal is that the mass
and velocity various for an ensemble of nano-diamond
\cite{Scala13}. In order to overcome this challenging, Ref.
\cite{Scala13} proposed to used Ramsay interferometry to observe the
interference.  With Ramsay interence, we can verify the matter wave
superpositions, as well as the quantum contextuality for macroscopic
quantum systems \cite{Asadian14}.  The basic setup of their poposal is
the same as Fig. \ref{fig:yin}. Here we assume that an angle $\theta$
between the vertical and the $z$ axis of the system. The Hamiltonian
of the system is
\begin{equation}\label{eq:Bose}
  H=\hbar D S_z^2+ \hbar \omega_r a_m^\dagger a_m - 2\hbar(\lambda S_z -\Delta\lambda) (a_m +a_m^\dagger),
\end{equation}
where $D=2.87$ GHz is the zero field splitting, and $\Delta \lambda =
\frac{1}{2}m g \cos\theta \sqrt{\frac{\hbar}{2m\omega_r}}$.  Let's
denote $\mu = 2(S_z\lambda - \Delta \lambda)/\hbar \omega_r$. The
initial state of resonator is a coherent state $|\beta\rangle$. The
spin dependent time evolution of the state is $|\beta(t,S_z)\rangle
|S_z\rangle$, where $S_z=0,\pm 1$. We get that
$$ \beta(t,S_z)\rangle = e^{-i(D-\omega_r \mu^2)t+i\mu^2 \sin (\omega_r t)} |(\beta-\mu) e^{-i\omega_r t} + \mu \rangle.
$$
A striking feature of this evolution is that the oscillator returns to
its original coherent state $\beta$ for any $\beta$ and $S_z$ at time
$t_0= 2\pi/\omega_r$.


As an example, we consider a initial separable state $|\beta\rangle
|S_z\rangle$. By using a strong microwave pulse $H_{mw} = \hbar \Omega
(|+1\rangle \langle 0| + |-1\rangle \langle 0| + \mathrm{h.c.}$, we
can change the state to $|\Psi(0)\rangle = \beta\rangle (|+1\rangle +
|-1\rangle)/\sqrt{2}$. The state after an oscillation period $t_0 =
2\pi/\omega_r$ becomes
\begin{equation}\label{eq:Ramsay}
  |\Psi (t_0) \rangle = |\beta \rangle \Big( \frac{|+1\rangle + e^{i\Delta \phi_g } |-1\rangle}{\sqrt{2}} \Big).
\end{equation}
where we have $\Delta \phi_g = 16\lambda \Delta \lambda t_0/(\hbar^2
\omega_r)$.  We apply microwave Hamiltonian $H_{mw}$ again to transfer
the spin population to $|0\rangle$.  After time $t_p = \pi /(2\sqrt{2}
\Omega)$, we get
$$ P_0 = \cos^2 \big( \frac{\Delta \phi_g}{2} \big),
$$
which can be used to measure the relative phase $\Delta \phi_g$. The
detection of $\Delta \phi_g$ is an evidence for the matter wave
superposition state.  Besides, as the phase $\Delta \phi_{g}$ is not
dependent to the initial state $\beta$, it is possible to be observed
when initial state is thermal.

\subsection{Applications as mass spectrometer}

As mentioned above, one of the most attractive features of the system
of optically levitated particles is its high mechanical quality
factor.  The trapped particle can coherently oscillator many times
(e.g., $10^{10}$ times) before its mechanical energy is damping out.
When this kind of almost-perfect mechanical oscillator couples to a
long-live quantum object, i.e. NV center electron spin, fantastic
quantum phenomena and novel application can be expected.

In Ref.~\cite{Zhao14}, we studied the quantum dynamics of a coupled system
of a quantum oscillator and a single spin, where the spin is under
dynamical decoupling control.  Dynamical decoupling control, as an
extended version of spin echo technique, has been proved to be an
efficient method to prolong the spin coherence time. In Ref.~\cite{Zhao14}
we predicted that, under dynamical decoupling control and coupled to a
high-quality mechanical oscillator, the spin coherence will exhibit a
series of periodic peaks, forming a comb structure (called time-comb
in Ref.~\cite{Zhao14}).  The peak width is calculated as
\begin{equation}
  \Delta_{{q}^{*}}\equiv \frac{2\sqrt{2}}{\gamma_{q^*}}
=\frac{T_0}{N\Lambda\sqrt{2n_{\text{th}}+1}}.
  \label{tQ expression}
\end{equation}

\begin{figure}[H]
  \centering
  \includegraphics[width=8cm]{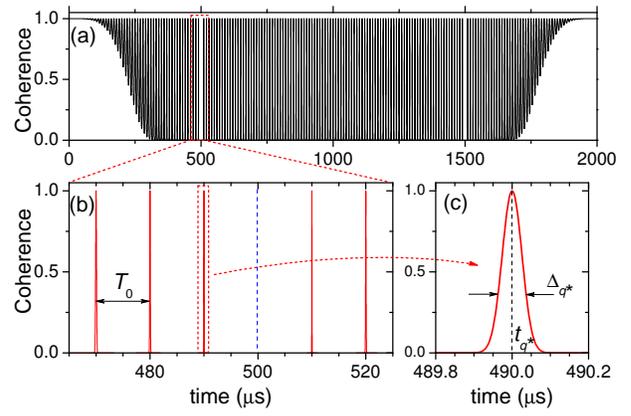}
  \caption{(a) The time-comb structure of qubit coherence
under $100$-pulse CPMG control. For $\omega_r t \gg 1$, the comb period
is synchronized with the oscillator period $T_0 = 2\pi/\omega_r$.
(b) Close-up of the coherence peaks. A missing peak at $\omega_rt =
N\pi$ is indicated by the blue dashed line. The peak width is decreasing when getting
close to the missing one.
(c) Close-up of the narrowest coherence
peak, which is centered at $t_{q^*}$ with width $\Delta_{q^*}$
(see text). The parameters used in this figure are oscillator frequency
$ 100$ kHz, coupling strengt $ \lambda =0.0001\omega_r$,
temperature $T=10$ K and $100$-pulse CPMG control.
}\label{fig:Zhao}
\end{figure}

Two features are notable in the above equation. First, the peak width
is inversely proportional to the square root of the thermal occupation
number $n_{\rm th}$ of the mechanical oscillator, which implies
narrower peaks in higher temperature.  Second, the peak width is
inversely proportional to the control pulse number $N$.  Since the
control pulse number $N$ is related to the total coherent evolution
time, and the later can be regarded to a kind of resource in
high-precision sensing, we expected that improving the dynamical
decoupling control pulse number can improve the measurement
sensitivity.

Based on the time-comb structure, we propose an ultra-sensitive mass
spectrometer using the coupled qubit-oscillator system.  Since the
peak position in the time axis is synchronized with oscillator period,
we propose to infer the oscillator period (and its mass assuming a
constant spring coefficient) by measuring the peak position.
Obviously according to the error propagation formula, the narrower the
peak is, the better sensitivity we will have.  The two features
mentioned above make the proposed mass spectrometer different to those
based on traditional measurement principles.  As the peak will be
narrower in higher temperature, we propose that the mass spectrometer
will have a counterintuitive temperature dependence of the sensitivity
(i.e., higher temperature helps improve the sensitivity).  This result
makes the room-temperature application of ultra-sensitive mass
spectrometer possible.  The second feature, as we showed in
Ref.~\cite{Zhao14}, implies an improved scaling of the measurement
sensitivity to the measurement resource.  The scaling relation goes
beyond the shot-noise scaling and Heisenberg scaling in the quantum
metrology theory.  The underlying physics which brings about this
improved scaling is under further studying.

\subsection{Experimental progresses}

\begin{figure}[H]
  \centering
  \includegraphics[width=7cm]{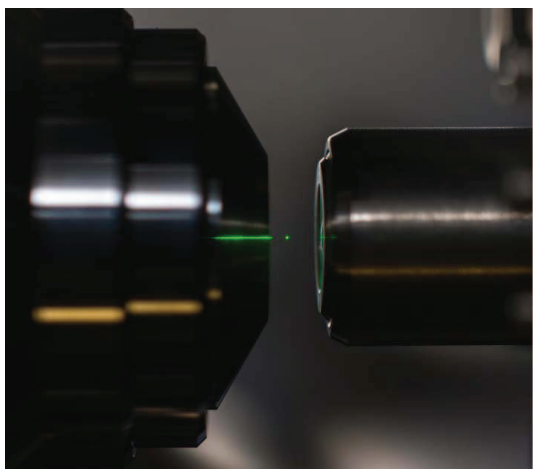}\\
  \caption{ A photograph of a $\sim 100$ diameter nanodiamond which is levitated
  by a optical tweezer ($1064$ nm). The green color is from an optical
  pump $532$ nm laser. The trapping lens is on the left, and the lens on the right is used to
collimate the light exiting the trap.
  Figure adopted from \cite{Neukirch14}. Image credit: J. Adam Fenster,
University of Rochester.}\label{fig:nanodiamond}
\end{figure}

Dielectric nanoparticles have been trapped and cooled optically,
as shown in Ref. \cite{Li11,Gieseler12,Kiesel13}.
Following the similar method, trapping nanodiamond in atmosphere
pressure has been realized experimentally
in 2013 \cite{Neukirch13}. The experimental setup is shown in Fig. \ref{fig:nanodiamond}.
They found the evidence of NV photoluminescence from a nanodiamond
in a free-space optical dipole trap. The photoluminescence rates
 are decreased with increasing trap laser power.
The neutral charge state (NV${}^0$) can be suppressed by continuous-wave
trap.

The charged nanodiamond was also levitated in a linear quadrupole
ion trap in air under ambient conditions \cite{Kuhlicke14}.
An electric trap has no trapping laser that quenches fluorescence emission. The surface
charge has effects on the emission rates of fluorescence spectrum of NV center. However, there is no
difference in the spectral properties was found in the experiment.
Up to now, there is no report on trapping nanodiamond in vacuum. This is the first problem that
should be overcome in future experiment.

\section{Conclusion and Outlook} \label{sec:Con}

In this paper we gave a brief review on the field of
(opto)nano-mechanical resonator coupling with NV centers electrons
spins. We discussed magnetic and strain induced phonon-spin coupling
mechanisms. We discussed how to cool the resonator to the ground state
and how to manipulate the phonon non-classical states. We also discuss
how to realize quantum information processing in this system. We
reviewed the progresses in optically trapped nano-diamond with
building in NV centers. We discussed how to observe macroscopic
quantum interference in the system. We also discussed how to used the
system as an ultra-sensitive mass spectrometer in room temperature.
Finally, we briefly summarized the experimental progresses.

Here we discuss some new directions and ideas in the field.  Up to
now, we only consider the first order magnetic gradient induced
coupling.  It would be interest to study the second order gradient
induced spin-resonator coupling.  This coupling mechanism may have
some new applications and features.  For example, it may be used to
entangle NV centers with different transition frequencies.  It may
also be used for cooling mechanical resonator which is out of
Lamb-Dicke regime.

Another new direction is to study the coupling between rotation degree
of freedom of trapped nano-diamond to the internal NV centers electron
spins. Geometric phase, even non-Abelian geometric phase may be
induced by mechanical rotation \cite{Maclaurin12,Kowarsky14}. The
rotation may also be cooled and manipulated by NV centers.

It would be interesting to investigate how to cool down the internal
temperature of the diamond which hosts the NV centers. As we know, the
decay rate of the NV centers electron spins is in verse proportional
to the temperature. Only under the low temperature $T<10$ K, the decay
rate can be decreased to around Hz. By considering the coupling
between NV center electron spins and the phonons in the diamond, it
would be actively cooled down the temperature of a nanodiamond trapped
with an optical tweezers \cite{Seletskiy10,Zhang13}. In this way, we
may extend the lifetime of NV centers electron spins around second
even at room temperature environment.

Before conclusion, we note that the proposals and schemes discussed in
this review should also be applied in other similar systems, such as
hybrid mechanical resonator with quantum dots \cite{Bell14}, $^28$Si
nanoparticles with donor spins \cite{Steger12}, or nanocrystals doped
with rare-earth ions \cite{Afzelius10}.

\no




\vspace*{2mm} \Acknowledgements{\bahao ZQY is funded by the NBRPC
  (973 Program) 2011CBA00300 (2011CBA00302), NNSFC NO. 11105136,
 NO. 61435007. NZ is supported by NKBRP
(973 Program) 2014CB848700 and NNSFC No. 11374032 and
No. 11121403. TL acknowledges the support from Purdue University
through the startup fund.}

\end{multicols}

\end{document}